\documentclass[twocolumn,aps,prl,showpacs,amsmath,amssymb,floatfix]{revtex4}
\usepackage{graphicx,subfigure}
\usepackage{bm,eucal}
\begin{document}
\title{Large-scale anisotropy in scalar turbulence}
\author{Antonio~Celani$^1$ and Agnese~Seminara$^{1,2}$} 
\affiliation{ 
$^1$ CNRS, INLN, 1361 Route des Lucioles, 06560 Valbonne, France \\
$^2$ Dipartimento di Fisica, Universit\`a degli Studi di Genova, 
and INFN Sezione di Genova, Via Dodecaneso 33, 16100 Genova, Italy} 
\date{\today}
\begin{abstract}
The effect of anisotropy on the statistics of a passive tracer transported
by a turbulent flow is investigated. We show that under broad conditions an
arbitrarily small amount of anisotropy propagates to the large scales 
where it eventually dominates the structure of the concentration field.
This result is obtained analytically in the framework of an exactly solvable
model and confirmed by numerical simulations of scalar transport in
two-dimensional turbulence.
\end{abstract}
\maketitle 

The emergence of large-scale anisotropy arising from small-scale sources
is a phenomenon that spans the most diverse fields of physics. 
For instance, the
microscopic anisotropy of crystals in mantle rocks in the Earth's interior  
is believed to induce large-scale seismic anisotropy \cite{seismic2},
and some small anisotropic perturbation in an early cosmological era,
evolving through gravitational collapse, is thought to be responsible for
the large-scale anisotropy of the cosmic microwave background radiation 
\cite{cosmic}. Conversely, in statistical physics, microscopic details 
such as lattice anisotropies may be wiped out by the dynamics 
allowing a recovery of symmetry and universality at large scales 
\cite{universality}.
In the theoretical and experimental analysis of turbulence much attention 
has been devoted to the anisotropy of the fine scales of fluid motion 
(see, e.g., \cite{BP} and references therein). 
Here, we take a different
viewpoint and investigate the effect of anisotropy on
the large-scale statistics of turbulence.
In this Letter we show how, unexpectedly, 
breaking rotational invariance by an arbitrarily small amount 
at a given scale induces a strong anisotropy on the large scales,
and symmetry is never restored.
 
We consider the evolution of a passive tracer described by a concentration 
field  $\theta({\bm x},t)$ and 
transported by a turbulent flow  ${\bm v}({\bm x},t)$
\begin{equation}
\label{eq:1}
\partial_t \theta+{\bm v}\cdot \bm\nabla \theta= \kappa \Delta \theta+f,
\end{equation}
where ${\bm v}$ is an incompressible, statistically homogeneous and
isotropic velocity field. The external driving $f$ is the source 
of scalar field fluctuations acting at a characteristic scale $l_f$.
The turbulent cascade toward small scales produces 
fine-scale structures of concentration that
are eventually smeared out by diffusion at scales $r_d \ll l_f$,
resulting in a statistically stationary state where input and dissipation
are in balance on average.  
The pumping mechanism can be chosen so as to introduce a certain degree
of anisotropy, that propagates across scales and may in principle pervade
the system.
However, the disordered motion of fluid particles induced by the
underlying turbulent, isotropic medium, might be sufficient to restore 
rotational invariance at scales far below or above $l_f$.
Indeed, this is the case at small scales $r \ll l_f$, where it can be shown
that the dominant contribution to the statistics of the scalar field $\theta$
is isotropic \cite{BP}.
At large scales $r \gg l_f$, since no upscale cascade
of scalar fluctuations occurs, {\em a fortiori} one would expect 
an essentially isotropic concentration field.
On the contrary, here we give theoretical
and numerical evidence that large-scale statistics is dominated by the 
anisotropic contribution under very broad conditions. 
We show that: 
({\it i\/}) The correlation function
$\langle \theta({\bm r},t) \theta({\bm 0},t) \rangle$
at scales $r\gg l_f$ is dominated by its anisotropic component 
decaying as a power law with an anomalous scaling exponent, 
as opposed to the exponential fall-off of the isotropic part.
This result is obtained analytically in the framework 
of the exactly solvable
Kraichnan model and its validity for realistic flows is
demonstrated by numerical simulations of passive scalar advection in
the inverse cascade of two-dimensional turbulence.
({\it ii\/}) Large-scale anisotropy manifests itself in the concentration
field with the appearance of "pearl necklace" structures
aligned with the preferential direction 
imposed by the microscopic anisotropy.
({\it iii\/}) The loss of isotropy at large scales can be interpreted as
a breakdown of equilibrium Gibbs statistics for the
anisotropic degrees of freedom;
({\it iv\/}) In the Lagrangian interpretation of passive scalar transport
the emergence of anisotropic power-law decay of correlation is 
associated to a long-lasting memory of the initial orientation
of particle pairs advected by the flow.

Let us first consider the Kraichnan model of passive scalar 
advection (see, e.g., \cite{FGV} for a thorough review), where ${\bm v}$ is 
a Gaussian, self-similar, incompressible, statistically
homogeneous and isotropic, white-in-time, $d-$dimensional 
velocity field. Its statistics is characterized by the correlation
$S_{\alpha\beta}({\bm r})\delta(t)=
\langle[v_{\alpha}({\bm r},t)-v_{\alpha}({\bm 0},t)]
[v_{\beta}({\bm r},0)-v_{\beta}({\bm 0},0)]\rangle=2 D
r^{\xi}[(d+\xi-1)\delta_{\alpha\beta}-\xi\, r_{\alpha}r_{\beta}/r^2] \delta(t)$.
The exponent $\xi$ measures the degree of roughness
of the velocity field and lies in the range $0<\xi<2$, the two extremes
corresponding to Brownian 
diffusion and smooth velocity, respectively.   
The assumption of $\delta$-correlation is of course far from being
realistic, yet it has the remarkable advantage of leading to 
closed equations for equal-time correlation functions.
In the following it will be sufficient to focus on 
the two-point correlation function $C({\bm r})=
\langle \theta({\bm r},t)\theta({\bm 0},t) \rangle$. In the
limit of vanishing diffusivity $\kappa \to 0$ and in the statistically
stationary state, $C$ obeys the 
partial differential equation ${\cal M} C({\bm r}) = - F({\bm r})$
where ${\cal M}= \frac{1}{2} 
S_{\alpha\beta}({\bm r}) \frac{\partial}{\partial r_\alpha}
  \frac{\partial}{\partial r_\beta}$. Here
$F$ is the correlation function of the Gaussian, white-in-time,
statistically stationary, homogeneous, anisotropic
 forcing $\langle f({\bm r},t) f({\bm 0},0)\rangle= F({\bm r})\delta(t)$.
At scales $r\lesssim l_f$ it equals  
the average input rate of scalar and then decays rapidly to zero, 
e.g. exponentially, as $r \gg l_f$.
By virtue of the statistical isotropy of the velocity field,
the operator ${\cal M}$ assumes a particularly simple form in radial
coordinates: ${\cal M}=D [ (d-1) r^{1-d}\partial_r r^{d-1+\xi} \partial_r +
(d+\xi-1) r^{\xi-2} {\cal L}^2] $ where ${\cal L}^2$ 
is the $d$-dimensional 
squared angular momentum operator. 
 It is then convenient to
decompose the correlation functions on a basis of eigenfunctions of
angular momentum ${\cal L}^2 Y_j = - j(j+d-2) Y_j $
with positive integer $j$. 
The short-hand notation $Y_j(\hat{\bm r})$, where $\hat{\bm r}={\bm r}/r$, 
does not account for
degeneracies and stands for the trigonometric functions in $d=2$ and the
spherical harmonics in $d=3$. 
Accordingly, we define the components of the
correlation functions in the $j$-th anisotropic sector as
$C({\bm r})=\sum_{j} C_{j}(r)Y_{j}(\hat{\bm r})$ and similarly for 
$F({\bm r})$, where $C_j$ and $F_j$ depend on $r=|{\bm r}|$ only. 
This yields a system of uncoupled 
differential equations in the radial variable for 
each anisotropic component $ {\cal M}_j C_{j}(r)= - F_{j}(r)$, where 
$ {\cal M}_j = D [ (d-1) r^{1-d}\frac{d}{dr} r^{d-1+\xi} \frac{d}{dr}- 
j(j+d-2)(d+\xi-1) r^{\xi-2} ] $, that can be solved
in each sector $j$. The resulting $C_j(r)$ is a linear combination of
a particular  solution determined by  $F_j(r)$, and
a homogeneous one $Z_j(r)$, a ``zero mode''. It is easy to see
that the former behaves as $r^{2-\xi +j}$ for 
$r\ll l_f$ (recall that $F_j \sim r^j$ at small $r$ if  $F$ is analytic
in the neighborhood of $r=0$) 
and that it must fall off exponentially fast for $r\gg l_f$ , as dictated 
by the decay of $F_j$.
The homogeneous solutions are
$Z^{\pm}_j(r)= r^{\zeta^{\pm}_j}$ with scaling exponents
$\zeta^{\pm}_j= \frac{1}{2}[-d+2-\xi\pm\sqrt{(d-2+\xi)^2+
4\frac{j(j+d-2)(d+\xi-1)}{d-1}}]$. 
The zero mode with positive scaling exponent
$\zeta^+_j$ appears at small scales  whereas the zero mode with
negative scaling exponent $\zeta^-_j$ 
is relevant in the range $r \gg l_f$.
In order to fully characterize the large-scale behavior of the
correlation function, it is necessary to identify the prefactor 
appearing in front of the homogeneous solution. This can be accomplished 
by writing the equation for $C_j$ in integral form:
$C_j(r)=\int_0^{\infty}G_j(r,\rho)F_j(\rho)d\rho$,
where $G_j(r,\rho)$ is the kernel of  
$-{\cal M}_j^{-1}$, i.e. the solution of 
${\cal M}_j G_j(r,\rho)=-\delta(r-\rho)$. The explicit form is
$G_j(r,\rho)=A(\rho)Z_+^j(r)Z_-^j(\rho)$ for $r<\rho$ and 
 $G_j(r,\rho)=A(\rho)Z_-^j(r)Z_+^j(\rho)$ for $r>\rho$, with
$A(\rho)=\rho^{d-1}/[D(\zeta^+_j-\zeta^-_j)]$.
Plugging this expression in the integral form for the correlation
function yields a large-scale behavior 
$C_j(r) \approx Q_j Z_j^-(r)$ + exponentially
decaying terms. The quantity 
$Q_j =\int_0^\infty A(\rho)Z_+^j(\rho) F_j(\rho)\,d\rho $ 
is of crucial importance and
plays the role of a ``charge'' in analogy with electrostatics \cite{FF05}.
In the isotropic sector $j=0$, it reduces to 
$Q_0=\frac{\Gamma(d/2)}{2 \pi^{d/2} D(d-2+\xi)} \int F({\bm r})\, d{\bm r}$.
When the isotropic charge $Q_0 \neq 0$, the leading behavior 
at large scales is isotropic, $C({\bm r})\sim C_0(r) \sim Q_0 r^{-d+2-\xi}$. 
The most interesting situation is when
$Q_0=0$, corresponding to the broad class of forcings
localized in wavenumber space ($Q_0 \propto \hat{F}({\bm k}={\bm 0})$).
In this event, there is
no power-law contribution from the isotropic zero mode and therefore 
the isotropic part of the correlation function is characterized by
an exponential decay at large $r$ \cite{FF05,CS05}. 
In the anisotropic sectors, it appears 
immediately that there is no reason to expect a null charge and
the generic situation is $Q_j \ne 0$ for $j>0$ (see \cite{example}).
As a result, {\em the large-scale correlation is dominated 
by the anisotropic contribution} arising from the zero-mode, 
$C({\bm r})\sim r^{\zeta^-_j} Y_j(\hat{\bm r})$,
that largely overweights the exponentially small isotropic part.
Among the various contributions arising from different sectors, the leading one
 corresponds to the lowest nonzero $j$ excited by the forcing, 
typically $j=2$ (odd $j$'s are switched on only by 
breaking reflection invariance). 

\begin{figure}
\includegraphics[width=7cm]{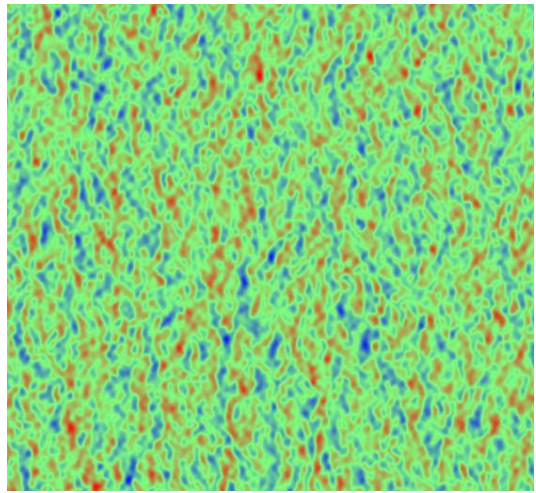}
\caption{Image of a scalar field for the Kraichnan model at 
$\xi=0$, $d=2$, $Q_0=0$, $Q_2 \neq 0$. 
The width of a single scalar patch is $\sim l_f$.
Forcing is preferentially acting in the vertical direction.
Here $C_0(r) \sim e^{-r^2/l_f^2}$ and $C_2(r)\sim r^{-2}$ at $r\gg l_f$. }
\label{fig:1}
\end{figure}

For the sake of illustration, we show in Fig.~\ref{fig:1} an
instance of a scalar field corresponding to the simple case
of Kraichnan advection by a very rough velocity ($\xi=0$), in $d=2$
and with $Q_0=0$.
Large-scale anisotropy manifests itself in the appearance of 
"pearl necklaces" made of like-sign scalar patches of size $\sim l_f$. 
These are aligned along the preferred direction of the forcing 
and extend for a length $\gg l_f$. 

It is worth pointing out the relationship between 
the appearance of anisotropic, 
anomalous scaling in the large-scale behavior of the scalar 
correlation function and equilibrium statistics. At large scales,
because of the absence of scalar flux, the system could be expected to
be in equilibrium and obey Gibbs statistics.
In physical space this corresponds to a concentration field
organized in independently distributed scalar patches of size $~l_f$. 
As recently shown in Refs.~\cite{FF05,CS05}, this is not true 
and substantial deviations are observed at the level of multipoint 
correlation functions already in the isotropic case. 
This departure has been traced back to the existence of nontrivial
zero-modes in that case as well. In the present case the 
breakdown of Gibbs statistics has an even more dramatic manifestation since it
occurs already for two-point correlation functions,
i.e. at the level of the spectral distribution of concentration. We now 
rephrase the previous findings
in terms of the averaged scalar spectral density, i.e. the Fourier transform
of the correlation function, $\hat{C}({\bm k})=
\langle |\hat\theta({\bm k},t)|^2 \rangle$, and its decomposition
in angular sectors in wavenumber space
$\sum_j \hat{C}_j(k)Y_j(\hat{\bm k}) $, where $\hat{\bm k}={\bm k}/k$.
For a correlation function decaying exponentially to zero at large $r$,
representative of large-scale equipartition in physical space,
the spectral density is analytic in a neighborhood of $k=0$.
In the series for  $\hat{C}({\bm k})$
the harmonic $Y_j(\hat{\bm k})$ appears only 
in the powers of ${\bm k}$ of order $\ge j$, yielding  
the long wavelength behavior $ \hat{C}^{(eq)}_j(k) \sim k^j$.
This defines the equipartition spectrum for generic anisotropic
fluctuations. However, because of the appearance of 
nontrivial zero modes in the anisotropic sectors the actual spectral density
contains also a contribution  
$\hat{C}^{(zero)}_j(k) \sim k^{-d-\zeta^-_j}$ that is responsible for 
the power-law decay of correlations in physical space with $j>0$.
For the Kolmogorov-Richardson value $\xi=4/3$ and the sector $j=2$ the 
anomalous spectrum always dominates the equipartition contribution
in spectral space as well ($-d-\zeta^-_{j=2} < 2$ for all 
$\xi<3/2$).

It is useful to reinterpret the results obtained so far 
within the framework of the Lagrangian 
interpretation of passive scalar transport. The correlation function
can be generically written as  $C({\bm r})=\int 
T({\bm \rho}|{\bm r})\, F({\bm \rho})\,  d{\bm\rho}$ 
where $T({\bm \rho}|{\bm r})\,d{\bm \rho}$ 
is the average time spent at a separation 
${\bm \rho}+d{\bm \rho}$ 
by a pair of particles that end their trajectories at a separation ${\bm r}$.
In the Kraichnan model $T$ is the kernel of the operator $-{\cal M}^{-1}$.
Since the action of the forcing is restricted to scales $\sim l_f$, 
the large-scale behavior of the correlation function 
is essentially dominated by the ensemble of trajectories
that have spent in the past a sufficiently long time at a short distance 
$\rho \lesssim l_f \ll r$. When $r/\rho$ tends to infinity $T$
becomes independent of $\rho$ and we obtain $T \sim r^{-d+2-\xi}$ 
\cite{BGK98}. The dependency on the final orientation 
$\hat{\bm r}$ is also lost.
This leads to the estimate $C(r) \sim Q_0 r^{-d+2-\xi}$. However, 
as noticed previously, when $Q_0=0$ the isotropic part of the
correlation function receives only exponentially small contributions 
from the forcing in the range $\rho \sim r \gg l_f$. 
Let us now turn our attention to the
anisotropic part of the correlation function. Projecting $C({\bm r})$ 
over $Y_j(\hat{\bm r})$ for $j>0$ amounts to give different weigths, 
positive and negative, 
to particle pairs oriented in different directions $\hat{\bm r}$.
Therefore $C_j$ can be interpreted as a difference of times spent
at  $\rho \lesssim l_f$ by differently oriented pairs. The first key point
is that the trajectories preserve a long-lasting memory of their initial 
orientation, with a slow power-law decay in $r$ that reflects in the
behavior of the correlation function.
Indeed, it can be shown that in the Kraichnan model 
$T({\bm \rho}|{\bm r})=\sum_j b_j r^{\zeta^-_j} 
\rho^{\zeta^+_j} Y_j(\hat{\bm r})Y_j(\hat{\bm \rho})$ for $r>\rho$.
Plugging this expression in the integral form  of the correlation
function gives the result $C_j(r)\sim Q_j r^{\zeta^-_j}$ as above.
Here emerges the second important point, i.e.
the dependence of $T$ on $\rho$: differently oriented trajectories 
sample the forcing unevenly in scales as $\rho \sim l_f$ 
 and this results in a nonvanishing charge $Q_j$ for $j>0$.
\begin{figure}
\includegraphics[width=8cm]{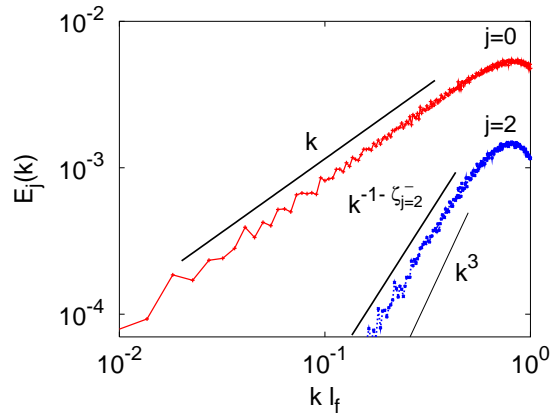}
\caption{Simulations of passive scalar advection by 
two-dimensional Navier-Stokes turbulence in the inverse cascade
(for details see \protect{\cite{CS05}}). Here are shown the
spectra $E_j(k)$ of concentration in the isotropic ($j=0$) 
and anisotropic ($j=2$) sectors. The spectral slopes are $1.0 \pm 0.05$
and $2.2\pm 0.1$, respectively ($\zeta^-_{j=2}=-3.2$).
The equilibrium spectra $k$ and $k^3$ are also shown
for comparison.}
\label{fig:2}
\end{figure}

\begin{figure}
\includegraphics[width=8cm]{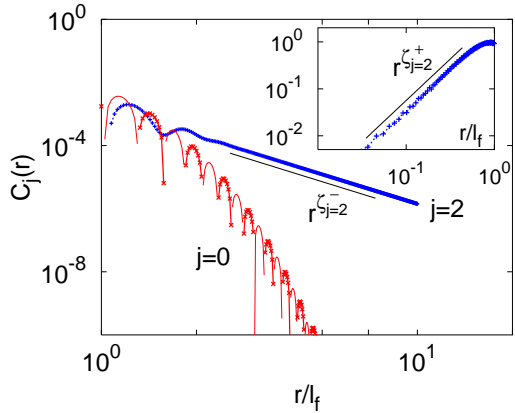} 
\caption{Main frame: 
Correlation functions $C_j(r)$ at large scales $r\gtrsim l_f$ 
for $j=0$ and $j=2$. 
Notice the exponential decay of the isotropic part as opposed to the 
 power-law behavior of the anisotropic component. 
In order to limit 
the statistical noise at large values of $r/l_f$ 
these two curves have been obtained by interpolating the spectra 
shown in Fig.~\protect\ref{fig:2} at small $k$ as power laws $k$ and 
$k^{-1-\zeta^-_{j=2}}$, respectively, and computing the correlation function 
from the integral $C_j(r)= \int_0^\infty J_j(k r) E_j(k) / k\, dk$.
Here $\zeta^-_{j=2} = -3.2$.
Inset: Correlation function $C_{j=2}(r)$ 
at small scales $r \lesssim l_f$.
The solid line is $r^{\zeta^+_{j=2}}$ with $\zeta^+_{j=2} = 1.8$. 
}
\label{fig:3}
\end{figure}

A remarkable advantage of the Lagrangian interpretation is that it 
does not make appeal to the special features of the Kraichnan model. This 
suggests that the same mechanisms are at work for realistic 
turbulent flows as well, and this expectation has been repeatedly 
confirmed for different aspects of passive scalar transport
\cite{CLMV00,CV01,CS05}. 
Here we show that anisotropy dominates the large-scale statistics
for real flows by showing the results of a numerical investigation of
passive scalar transport in the inverse cascade 
of two-dimensional Navier-Stokes turbulence.
This flow has been studied in great detail, both experimentally
in fast flowing soap films \cite{soap} 
and in shallow layers of electromagnetically driven electrolyte 
solutions \cite{electr-sol,RE05}, 
and numerically \cite{2D1,2D2,BBCF06}. 
The velocity field ${\bm v}$ is statistically homogeneous, isotropic, and 
scale-invariant with exponent $h=1/3$ ($\delta_r v \sim r^h$) 
in the range $l^v_f\lesssim r \lesssim L_v$, where 
$l^v_f$ denotes the kinetic energy injection length
and $L_v$ the velocity integral scale. 
The scalar field is governed by Eq.~(\ref{eq:1}) driven by 
a homogeneous, anisotropic, Gaussian, $\delta$-correlated driving 
$f$ that excites the sectors $j=0$ and $j=2$ and satisfies 
the condition of null isotropic charge (see \cite{example}). 
The various lengthscales  are ordered as follows: 
$l^v_f \ll r_d \ll l_f \ll L_v $.
In Fig.~\ref{fig:2} we show the spectral content of scalar fluctuations
$E_j(k)= \pi^{-1} k \int_0^{2\pi}\cos(j \phi_k)\, \hat{C}({\bm k})\,d\phi_k = 
k \hat{C}_j(k)$
at $k l_f < 1$, i.e. at large scales. The isotropic spectrum ($j=0$)
agrees very well with the Gibbs equilibrium distribution, 
$E_{j=0}(k) \propto k$,  and corresponds to
exponentially decreasing isotropic correlation at large scales
(see the main frame of Fig.~\ref{fig:3})
in agreement with the theoretical arguments presented above.
The anisotropic spectrum shows a power-law behavior
$E_{j=2}(k) \sim k^{2.2 \pm 0.1}$ 
definitely dominating over the equilibrium spectrum
$E^{eq}_{j=2}(k)\sim k^3$. 
In physical space, this translates in a power-law decay 
of the anisotropic correlation function $C_{j=2}(r)$, as shown in
the main frame of  Fig.~\ref{fig:3}, and leads to the estimate
$\zeta^-_{j=2} \simeq -3.2 \pm 0.1$. Therefore, the correlation
function at large scales is dominated by the anisotropic power-law decay
for 2D Navier-Stokes advection as well. 
Finally, we notice that for incompressible, 
time-reversible, self-similar flows 
the two zero-mode exponents are 
conjugated by the dimensional relation  
$\zeta^+_j+\zeta^-_j=-d+1-h$
(within the Kraichnan model $1-h=2-\xi$, due to the
$\delta$-correlation in time).  
In the inset of Fig.~\ref{fig:3} we show the behavior of the correlation
function $C_{j=2}(r)\sim r^{\zeta^+_j}$ at small scales $r\ll l_f$ 
that indeed displays an exponent $\zeta^+_j=1.8 \pm 0.1$ compatible with the
previous relation \cite{revers}.

In summary, we have shown that microscopic anisotropies introduced
by the forcing have a dramatic imprint on the large-scale statistics
of passive scalar turbulence.
From this result new questions arise naturally,
the most intriguing one being whether the 
large scales of hydrodynamic turbulence
show such striking properties as well.
Further theoretical, experimental and numerical effort in 
this direction is needed to elucidate this point.

This work has been partially supported by the EU contract HPRN-CT-2002-00300  and by CNRS PICS 3057.


\begin{thebibliography}{99}
\bibitem{example}
A simple example is given by the following choice of anisotropic driving
in $d=2$: the spectral density of the forcing
$\hat{F}(k,\phi_k)=\pi^{-1} k^2 l_f^4 \exp(-k^2 l_f^2/2) \sin^2\phi_k$
results in a forcing correlation function that
excites the two sectors $j=0,2$ only: 
$F(r,\phi)\!=\!F_0(r)+F_2(r)\cos(2\phi)$, where 
$F_0(r)\!=\!2 [1-r^2/(2l_f^2)]\exp[-r^2/(2 l_f^2)]$
and $F_2(r)\!=\!-(r/l_f)^2 \exp[-r^2/(2 l_f^2)]$.
The resulting charges are
$Q_0=(D\xi)^{-1} \int_0^\infty f_0(r)r\,dr \propto\int_0^{\infty} r 
[1-r^2/(2l_f^2)] \exp[-r^2/(2l_f^2)]
dr = 0$ and $Q_2= 
(\sqrt{2} l_f)^{2+\zeta^+_2}
\Gamma(2+\zeta^+_2/2)/[D(\zeta^+_2-\zeta^-_2)] \ne 0$.
\bibitem{revers}
Notice that, strictly speaking, the assumption of time-reversibility 
for the inverse cascade is incorrect, due to the presence of a kinetic 
energy flux. However, the almost-Gaussian statistics of velocity (see, e.g., 
Ref.~\cite{2D2}) suggests that it is a good approximation.
\end{thebibliography}
\end{document}